\documentclass[journal]{IEEEtran}

\usepackage{graphicx}
\usepackage{cite}
\usepackage{soul}
\usepackage{framed}
\usepackage{epsfig,mathptmx,times,amsmath,amssymb,color,flushend,gensymb,subfigure,fancyhdr,subfigure}
\usepackage{framed}
\definecolor{shadecolor}{rgb}{1,0.8,0.3}

\ifCLASSINFOpdf
\else
\fi
\allowdisplaybreaks 
\begin{document}

\title{Learning in Centralized Nonlinear Model Predictive Control: Application to an Autonomous Tractor-Trailer System}

\author{Erkan~Kayacan,~\IEEEmembership{Student Member, IEEE,} ~~Erdal~Kayacan,~\IEEEmembership{Senior Member, IEEE,} ~~Herman~Ramon~\IEEEmembership{}~and~Wouter~Saeys~\IEEEmembership{}% <-this % stops a space
\thanks{E. Kayacan, H. Ramon and W. Saeys are with the Division of Mechatronics, Biostatistics and Sensors, Department of Biosystems, University of Leuven (KU Leuven), Kasteelpark Arenberg 30, B-3001 Leuven, Belgium.
e-mail: {\tt\small \{erkan.kayacan, herman.ramon, wouter.saeys\}@biw.kuleuven.be }}
\thanks{E. Kayacan is with the School of Mechanical and Aerospace Engineering, Nanyang Technological University, 639798, Singapore.
e-mail: {\tt\small erdal@ntu.edu.sg }}
}

\markboth{\textbf{PREPRINT VERSION:} IEEE TRANSACTIONS ON CONTROL SYSTEM TECHNOLOGY, vol. 23, no. 1, pp. 197-205, Jan. 2015.}
{Shell \MakeLowercase{\textit{et al.}}: Bare Demo of IEEEtran.cls for Journals}
\maketitle

\begin{abstract}
One of the most critical tasks in tractor operation is the accurate steering during field operations, e.g. accurate trajectory following during mechanical weeding or spraying, to avoid damaging the crop or planting when there is no crop yet. In order to automate the trajectory following problem of an autonomous tractor-trailer system and also increase its steering accuracy, a nonlinear model predictive control approach has been proposed in this paper. For the state and parameter estimation, moving horizon estimation has been chosen since it considers the state and the parameter estimation within the same problem and also constraints both on inputs and states can be incorporated. The experimental results show the accuracy and the efficiency of the proposed control scheme in which the mean values of the Euclidean error for the tractor and the trailer respectively are $6.44$ cm and $3.61$ cm for a straight line trajectory and $49.78$ cm and $41.52$ cm for a curved line trajectory.
\end{abstract}


\IEEEpeerreviewmaketitle

\section{Introduction}

\IEEEPARstart{W}{hereas} human population has been increasing remarkably, total size of arable lands on Earth has been decreasing. According to the Food and Agriculture Organization of the United Nations (FAO), the population of the world will reach 9.1 billion by 2050, and in order to feed this larger population, food production will have to be nearly doubled \cite{fao}. In the future, agricultural production machines with higher efficiencies and better precisions, which can be realized by the automatization of them, will be demanded. The reason behind automating agricultural production machines is the poor performance of manual operators especially for multi-tasking operations. It is to be noted that a manual driver has not only to drive the agricultural vehicle with a high accuracy but also he/she has to adjust several parameters, such as trailer position and additional parameters of some specific agricultural operation equipments. In such cases, an advanced control algorithm for the trajectory tracking of a production machine, e.g. a tractor-trailer system, is more than welcome to increase the efficiency of the human operator. Such an advanced control algorithm will make the operator free from the challenging navigation task and allow him/her to fully focus on the work to be done. Furthermore, the accuracy of the trajectory following will not decrease under challenging working conditions such as during the night or long working hours. This paper considers the trajectory tracking problem of an autonomous tractor-trailer system with various sensors and actuators as an agricultural production machine.

Despite a wide variety of theoretical studies on the trajectory tracking problem for autonomous tractor-trailer systems, there is still a long way to go to deal with the challenges for real-time implementation. First, the sensors and the electro-hydraulic valves to automate the tractor-trailer system are highly nonlinear, having dead-bands and hysteresis characteristics. Second, the inputs to the actuators and the states of the system should be constrained for environmental and safety reasons. Third, several states and parameters of the system cannot be measured, and thus have to be estimated online during field operation.

%In this paper, a nonlinear  model predictive control (NMPC) algorithm has been preferred to deal with the trajectory tracking problem of an autonomous tractor-trailer system to cope with all the challenges mentioned above.

As an agricultural production machine, a tractor-trailer system is a complex mechatronic system which consist of several subsystems that interact with each other as a result of energy and information flows. For instance,  the hydrostatic drive line, the steering system of the tractor and the steering system of the trailer are fed by the same hydraulic pump coupled to the diesel engine. In other words, whenever an input signal is applied to one of the mentioned three subsystems, the other two subsystems are also affected. One could use a decentralized control approach, \emph{e.g.} decentralized model predictive control (MPC), to deal with the mentioned control problem. However, in most of the cases, these controllers are not aware of the interactions with other subsystems, and they will exhibit selfish behavior leading to suboptimal performance of the global system \cite{Scattolini2009}. An alternative solution is the use of a centralized control approach, \emph{e.g.} centralized MPC. However, the main criticism on the centralized control approach in literature is that the centralized control of large scale systems using a plant-wide model may not be computationally feasible since the optimization process of a multi-input-multi-output control system is computationally too demanding \cite{Liu2010,Lynch}. As a possible solution for the aforementioned problem, in order to be able to solve the optimization problems in NMPC, the \emph{ACADO} code generation tool \cite{ACADO} has been used which is capable of giving feedback times in milliseconds range (around 2-3 ms). Moreover, instead of using a dynamic model, a tricycle kinematic model has been used to define the mathematical behaviour of the tractor-trailer system. All the implementation preferences above allow us to use a centralized NMPC (CeNMPC) method for the real time system at hand.

%In order to cope with the challenges mentioned above, a  real time kinematic (RTK) global positioning system (GPS) has been used to determine the global position of the tractor and the trailer, and the preferred GPS is able to provide new measurements in every 200ms. That is why, the overall control structure of the system has been run at 50Hz.

While MPC has always been a popular technique in the process industry, the processes are staying around fixed operating-points in most of the applications in the process industry. This allows linearization of the process model and thus the application of linear MPC. However, since the autonomous tractor-trailer system has time-varying set points, highly nonlinear actuators and several disturbances,  the idea of local linearization is excluded. Therefore, more dedicated research towards NMPC is necessary. Moreover, there are more constraints both on the inputs and the states of the system which have to be satisfied. As an extension of conventional linear MPC, the NMPC approach is a proper candidate to meet the mentioned demands.

Since all the states describing the dynamics of the system to be controlled must be fed to NMPC, they have to be directly measured or estimated. In practical applications, since there exist always some unmeasurable states, it is generally necessary to estimate some states or unknown model parameters online when working with NMPC. A common practical solution in literature for the state and parameter estimation is the use of the extended Kalman Filter (EKF) approach. However, the main disadvantage of the EKF approach is that this method is not capable of dealing with the constraints on states or parameters as they are often imposed. In this paper, an alternative method, nonlinear moving horizon estimation (NMHE) has been preferred since it treats the state and the parameter estimation within the same problem and also constraints can be incorporated \cite{TomKraus}.

Various implementation examples to control a tractor with/without trailer system have been presented in literature. In order to follow straight lines, model reference adaptive control was proposed for the control of a tractor configured with different trailers in \cite{Derrick}, and a linear quadratic regulator was used to control a tractor-trailer system in \cite{karkeejournal}. Both controllers are designed based on dynamic models; however, since these dynamic models are derived with a small steering angle assumption, they would give poor performance for curvilinear trajectory tracking. For curvilinear trajectories, NMPC was proposed for the control of a tractor-trailer system in \cite{Backman2012}. EKF was used to estimate the yaw angles of the tractor and trailer. However, side-slips of the tractor and trailer were not estimated and not considered for the control of system. The states and parameters of a tractor including wheel slip and side-slip were estimated with NMHE and fed to an NMPC in \cite{TomKraus}. This concept will be further extended to the tractor-trailer system.

\emph{Contribution of this paper: } In this paper, a fast CeNMPC based on \emph{ACADO} code generation tool \cite{ACADO, Diehl}  is combined with NMHE to obtain accurate trajectory tracking of an autonomous tractor-trailer system under unknown and variable soil conditions. The developed framework has been tested in real-time.

This paper is organized as follows: The kinematic model of the system is presented in Section \ref{SystemModel}. The basics of the implemented CeNMPC approach and the learning process by using NMHE are explained in Section \ref{nmpcmhe}.  The experimental set-up and the experimental results are described in Section \ref{realtime}. Finally, some conclusions are drawn from this study in Section \ref{Conc}.
%%%%%%%%%%%%%%%%%%%%%%%%%%%%%%%%%%%%%%%%%%%%%%%%%%%%%%%%%%%%%%%%%%%%%%%%%%%%%%%%%%%%%%%%%%%%%%%%%%%%%%%%%%%%%%%%%%%%%%%%%%%%%%%%%%%%%%%%%%%%%%%%%%%%%%%%%%%%%%%%

\section{System Model}\label{SystemModel}
\begin{figure}[b!]
\centering
  \includegraphics[width=2.5in]{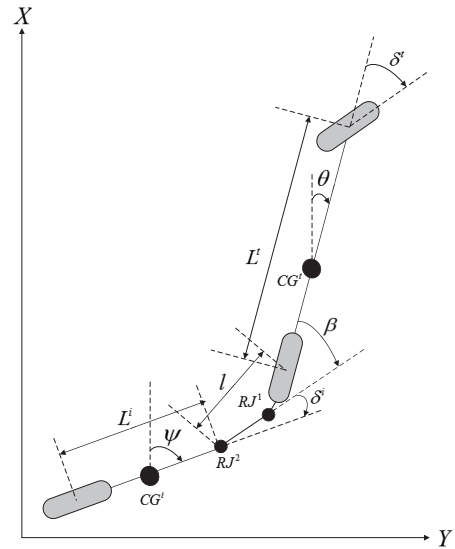}\\
  \caption{Schematic illustration of tricycle model for an autonomous tractor-trailer system}
  \label{kinematic}
\end{figure}

The schematic diagram of an autonomous tractor-trailer system is presented in Fig. \ref{kinematic}.

The model for the autonomous tractor-trailer system is \emph{a kinematic model} neglecting the dynamic force balances in the equations of motion. A dynamic model would, of course, represent the system behaviour with a better accuracy, but then a system identification and multibody modelling techniques would be needed for obtaining an accurate dynamic model of the system. Moreover, a dynamic model would increase the computational burden in the optimization process in CeNMPC. Thus, an extension of a simpler well-known tricycle kinematic model in \cite{karkeejournal,karkeephd} has been used for the CeNMPC design in this paper. The extensions are the additional three slip parameters ($\mu$, $\kappa$ and $\eta$) and the definition of the yaw angle difference between the tractor and the trailer by using two angle measurements ($\delta^{i}$ and $\beta$) instead of one angle measurement. These slip parameters stand for making the model adaptive.

The equations of motion of the system to be controlled are as follows:
\small
\begin{eqnarray}\label{kinematicmodel}
\left[
  \begin{array}{c}
  \dot{x}^{t}  \\
  \dot{y}^{t}  \\
  \dot{\theta} \\
  \dot{x}^{i}  \\
  \dot{y}^{i}  \\
  \dot{\psi}   \\
  \end{array}
  \right]
=
\left[
  \begin{array}{c}
  \mu v \cos{(\theta)} \\
 \mu v \sin{(\theta)}  \\
\frac{\mu v \tan{ (\kappa \delta^{t}) } }{L^t} \\
 \mu v \cos{(\psi)}  \\
 \mu v \sin{(\psi)}  \\
\frac{\mu v }{L^i}\big(\sin{(\eta \delta^{i} + \beta)} + \frac{l }{L^t} \tan{ (\kappa \delta^{t})} \cos{(\eta \delta^{i} + \beta)} \big)  \\
  \end{array}
  \right]
\end{eqnarray}
\normalsize
where ${x}^{t}$ and $y^{t}$ represent the position of the tractor, $\theta$ is the yaw angle of the tractor, ${x}^{i}$ and $y^{i}$ represent the position of the trailer, $\psi$ is the yaw angle of the trailer, $v$ is the longitudinal speed of the system. Since the tractor and trailer rigid bodies are linked by two revolute joints at a hitch point, the tractor and the trailer longitudinal velocities are coupled to each other.  The steering angle of the front wheel of the tractor is represented by $\delta^{t}$, $\beta$ is the hitch point angle between the tractor and the drawbar at $RJ^{1}$; $\delta^{i}$ is the steering angle between the trailer and the drawbar at $RJ^{2}$; $\mu$, $\kappa$ and $\eta$ are the slip coefficients for the wheel slip of the tractor, side-slip for the tractor and side-slip for the trailer, respectively. It is to be noted that the slip parameters can only get values between zero and one. While a wheel slip of one indicates that the wheel and tractor velocities are the same, a ratio of zero indicates that the tire is not rotating and the wheels are skidding on the surface, i.e., the tractor is no longer steerable \cite{Kayacan2009,Topalov2011}. A similar physical explanation can be made for the other two slip parameters.

The physical parameters that can be directly measured are as follows: $L^t=1.4m$, $L^i=1.3m$, and $l=1.1m$.  The parameters $L^t$, $L^i$, and $l$ represent the distance between the front axle of the tractor and the rear axle of the tractor, the distance between $RJ^2$ and the rear axle of the trailer, and the distance between $RJ^1$ and $RJ^2$, respectively.

%%%%%%%%%%%%%%%%%%%%%%%%%%%%%%%%%%%%%%%%%%%%%%%%%%%%%%%%%%%%%%%%%%%%%%%%%%%%%%%%%%%%%%%%%%%%%%%%%%%%%%%%%%%%%%%%%%%%%%%%%%%%%%%%%%%%%%%%%%%%%%%%%%%%%%%%%%%%%%%%
\section{Nonlinear Model Predictive Control and Nonlinear Moving Horizon Estimation Framework} \label{nmpcmhe}
\subsection{Centralized Nonlinear Model Predictive Control Formulation}

In single-input-single-output control scheme, a state is generally tried to be followed a constant value or a time-varying value by using one control variable. However, in multi-input-multi-output control scheme, since multi states are tried to be controlled by using multi control variables, it is more challenging work than the previous one. In this case, MPC approach might be capable off controlling wide-process model. These wide-process models are generally nonlinear and not convenient to be used in linear control techniques. This results in a necessity of the combination of a nonlinear model and an MPC which is referred to NMPC.

Since NMPCs have the ability to predict the future system behavior based-on a system model, they are designed based-on a model of the system to be controlled \cite{Maciejowski}. The objective function consists of a function of the states, the outputs and the inputs of the system. The control action is calculated by minimizing the cost function subject to the predicted behavior of the model and the system constraints. The states and the outputs are predicted over a given prediction horizon. The main equality constraint is the system model, and the inequality constraints are the state constraints, the output and the input constraints (actuators limits).

In this paper, the NMPC formulation of the following form has been considered:

\begin{equation}
 \begin{aligned}
 & \underset{x(.), u(.)}{\text{min}}
 & & \int^{t_k+t_h}_{t_k} \big(\| x_{r} (t) - x (t) \|^{2}_{Q} + \| u_{r} (t) - u (t) \|^{2}_{R} \big) dt  \\ &
 && + \| x_{r} (t_k+t_h) - x (t_k+t_h) \|^{2}_{S} \\
 & \text{subject to}
& & x(t_k)= \hat{x}(t_k) \\
 && & \dot{x}(t) = f \big(x(t),u(t),p \big) \\
 &&& x_{min} \leq x(t) \leq x_{max} \\
 &&& u_{min} \leq u(t) \leq u_{max} \;\; \text{for all} \;\; t \in [t_k, t_k+t_h]\\
  \end{aligned}
  \label{nmpc}
\end{equation}
where $x_{r}$ and $u_{r}$ are the references for the states and the inputs, $x$ and $u$ are the states and the inputs, $t_k$ stands for the current time, $t_h$ represents both the prediction horizon and control horizon, $p$ represents the model parameters. Upper and lower bounds on the states and the inputs are represented by $x_{min}$, $x_{max}$, $u_{min}$ and $u_{max}$. Moreover, the matrices $Q$, $R$ and $S$ in the objective function are weighing matrices being symmetric and positive definite. These $Q$, $R$ and $S$ stand for the deviations of the model response to reference trajectories and the future value of the deviations of the model response raised by the controlled variables at the end of the given horizon $t_{h}$, respectively. The last part of the objective function is referred to as \emph{terminal cost} and stated for stability reasons in NMPC \cite{Mayne}.

\subsubsection{Implementation of CeNMPC}

The CeNMPC problem is solved at each sampling time with the following constraints on the inputs, which are the steering angle references for the tractor and the trailer:
\begin{eqnarray}\label{constraints}
-35 \degree  \leq & \delta^{t}(t) & \leq 35 \degree \nonumber \\
-25 \degree  \leq & \delta^{i}(t) & \leq  25 \degree
\end{eqnarray}

The references for the positions and the inputs of the tractor and trailer are changed online as follows:
\begin{eqnarray}\label{}
x_{r} & = & (x^t_{r},y^t_{r}, \theta_{r}, x^i_{r},y^i_{r}, \psi_{r})^T \nonumber \\
u_{r} & = & (\delta^{t}_{r},\delta^{i}_{r})^T
\end{eqnarray}

The input references are the recent measured steering angle of the front wheel of the tractor and the steering angle of the trailer. They are used in the objective function to provide a possibility to penalize the variations of the inputs from timestep to timestep. Moreover, the weighting matrices $Q$, $R$ and $S$ are defined as follows:
\begin{eqnarray}\label{weightingmatricesQRS}
Q & = & diag(0.5,0.5,0,0.005,0.005,0) \nonumber \\
R & = & diag(5,0.05) \nonumber \\
S & = & diag(5,5,0,0.05,0.05,0)
\end{eqnarray}
As can be seen from \eqref{weightingmatricesQRS}, when the equations of motion \eqref{kinematicmodel} are considered, two remarks can be made:
\begin{enumerate}
  \item Although the input to the tractor affects the trailer, the states of the tractor do not  have such an effect on the trailer.
  \item Neither the input nor the states of the trailer have an effect on the tractor.
\end{enumerate}
For this reason, the input of the trailer only tries the minimize the position errors of the trailer on x- and y-axes. However, the input of the tractor tries to minimize the position errors of the tractor and trailer on x- and y-axes.

During the experiments, it has been observed the following relation between the weighting matrices Q and R in \eqref{weightingmatricesQRS}: When the coefficients for the positions for the tractor and trailer in the weighting matrix Q have been set to equal to each other or the ones for the trailer has been chosen larger than the ones for the tractor, an oscillatory behavior has been observed due to the fact that the actuator for the trailer is slower rather than the actuator for the tractor. As a result, the coefficients for the trailer in the weighting matrix Q in \eqref{weightingmatricesQRS} has been chosen smaller than the ones for the tractor. Thus, the input of the tractor tries to minimize the position errors of the tractor rather than the position errors of the trailer. If we had had a faster actuator for the trailer, more balanced selection of Q in \eqref{weightingmatricesQRS} could have been done. Moreover, another reason for the oscillatory behavior might be the reason that our kinematic model from the tractor steering input to the implement yaw angle may not be accurate enough. Our solution in the paper compensate both errors mentioned above, because our method steers the tractor accurately on the desired trajectory and then steers the implement with respect to this.

Moreover, the weighting matrix $R$ for the inputs is chosen large in order to get well damped closed-loop behaviour. The reason is that since the tractor-trailer system dynamics are slow, it cannot give a fast response. Besides, the weighting matrix $S$ for the future value of the states error at the end of the predictive horizon is set to $10$ times bigger than the weighting matrix $Q$ for the states error. Thus, deviations of the predicted values at the end of the horizon from their reference are penalized $10$ times more in the CeNMPC cost function than the previous points. This is because the error value at the end of the prediction horizon is the most important for the control algorithm. The prediction horizon and control horizon $t_h$ is set to $3$ seconds which was found by trail-and-error method.

\subsection{Nonlinear Moving Horizon Estimation}

EKF is the most commonly used method for the state and parameter estimation of nonlinear systems. In this method, uncertainties on the measurements are assumed to be gaussian and constraints neither on the inputs nor on the states are incorporated \cite{Daum,Haseltine}. Unlike EKF, as an optimization-based estimation method, an NMHE treats the state and the parameter estimation within the same problem and also constraints can be incorporated \cite{Rao2000,Rao2003,Kuhl2011,TomKraus}. The constraints play an important role in the autonomous tractor-trailer system. For instance, the slip coefficients in  \eqref{kinematicmodel} cannot be bigger than $1$.

The NMHE problem is formulated as follows:
\begin{equation}
 \begin{aligned}
 & \underset{x(.),p,u(.)}{\text{min}}
 & & \int^{t_k}_{t_k-t_h} \big(\| y_m (t) - y (t) \|^{2}_{V_y} + \| u_m (t) - u (t) \|^{2}_{V_u} \big) dt  \\ &
 && + \left\|
  \begin{array}{c}
    \hat{x} (t_k-t_h) -x (t_k-t_h)  \\
    \hat{p} - p
 \end{array}
 \right\| ^{2}_{V_s} \\
 & \text{subject to}
 &&  \dot{x}(t) = f \big(x(t),u(t),p \big) \\
 &&&  y(t) = h \big(x(t),u(t),p \big) \\
 &&& x_{min} \leq x(t) \leq x_{max} \\
 &&& p_{min} \leq p \leq p_{max} \;\; \text{for all} \;\; t \in [t_k - t_h ,t_k]\\
  \end{aligned}
    \label{mhe}
\end{equation}
where $y_m$ and $u_m$ are the measured outputs and inputs, respectively. Deviations of the first states in the moving horizon window and the parameters from priori estimates $\hat{x}$ and $\hat{p}$ are penalized by a symmetric positive definite matrix $V_s$. Moreover, deviations of the system outputs and the measured outputs and deviations of the system inputs and the measured inputs are penalized by symmetric positive definite matrices $V_y$ and $V_u$, respectively \cite{Ferreau}. Upper and lower bounds on the model parameters are represented by parameters $p_{min}$ and $p_{max}$, respectively.

The last term in the objective function in \eqref{mhe} is known as the arrival cost. The reference estimated values $\hat{x} (t_k-t_h)$ and $\hat{p}$ are taken from the solution of NMHE at the previous estimation instant. In this paper, the arrival cost matrix $V_s$ has been chosen as a so-called smoothed EKF-update based on sensitivity information obtained while solving the previous NMHE problem \cite{Robertson}.

The arrival cost in the NMHE formulation stands for the information before the beginning of the estimation horizon $t_k - t_h$. Theoretically, the cost function for all the measurements in the past can be summarized as follows:
\begin{equation}\label{cf_arrival}
\underset{x(.),p,u(.)}{\text{min}}
\int^{t_k-t_h}_{0} \big(\| y_m (t) - y (t) \|^{2}_{V_y} + \| u_m (t) - u (t) \|^{2}_{V_u} \big) dt  \\
\end{equation}
It is to be noted that the cost function in \eqref{cf_arrival} is replaced with the last term in \eqref{mhe}. Thus, this objective function becomes an NMHE formulation with the estimation horizon $t_k$. As can be seen from the formulation, when the estimation horizon $t_k$ goes to infinity, a solution for the NMHE problem will no longer exist. In order to avoid this problem, the estimation horizon length must be kept constant, and the arrival cost is approximated with a smoothed EKF update. This results in an EKF running with the measurements at time $t_k$ in order to update Kalman estimates $\hat{x}, \hat{p}$ and the inverted Kalman covariance $V_s$ for each NMHE iteration \cite{Robertson}. Thus, only initial values for $\hat{x}, \hat{p}$ and $V_{s} ^{-1}$ must be available with covariance matrices describing the error distributions for measurement and process noise. A square-root variant of the EKF update formulas has been applied in this study \cite{Kuhl2011}.

It is necessary that the arrival cost is bounded. Otherwise, if the inverted Kalman covariance $V_s$ is too heavy, the arrival cost can go to infinity. For this reason, the contributions of the past measurements to the inverted Kalman covariance $V_s$ are downweighted by a process noise covariance matrix \cite{TomKraus}.

\subsubsection{Implementation of NMHE}
Some states of the autonomous tractor-trailer system cannot be measured. Even if states can be measured directly, the obtained measurements contain delay and noise. Especially, in the case of the GPS measurement values are regularly missing due to either poor satellite connection or poor 3G connection.. In order to estimate the unmeasurable states or parameters, the NMHE method is used. Since only one GPS antenna is mounted on the tractor and one GPS antenna on the trailer, the yaw angles of the tractor and the trailer cannot be measured. It is to be noted that the yaw angles of the tractor and the trailer are crucial for the trajectory tracking control.

The inputs to the NMHE algorithm are the position of the tractor, the longitudinal velocity values coming from the encoders mounted on the rear wheels of the tractor and the steering angle values coming from the potentiometer in the front wheels of the tractor, the position of the trailer and the steering angle values coming from the inductive sensors on the trailer. The outputs of NMHE are the positions of the tractor and the trailer in the x- and y-coordinate system, the yaw angles for both the tractor and the trailer, the slip coefficients and the longitudinal speed. The estimated values are then used by the CeNMPC to calculate the input signals to be applied to the system.

The NMHE problem is solved at each sampling time with the following constraints on the parameters:
\begin{eqnarray}\label{constraints2}
0.25  \leq & \mu & \leq 1 \nonumber \\
0.25  \leq & \kappa & \leq 1 \nonumber \\
0.25  \leq & \eta & \leq 1
\end{eqnarray}
Even on an icy road, the slip parameters are expected to be around $0.2$. Thus, the lower limit is chosen to be equal to $0.25$ for an agricultural operation.

The standard  deviations of the measurements have been set to $\sigma_{x^t} = \sigma_{y^t} = \sigma_{x^i} = \sigma_{y^i} = 0.03$ m, $\sigma_{\beta} = 0.0175$ rad, $\sigma_{v} = 0.1$ m/s, $\sigma_{\delta^{t}} = 0.0175$ rad and $\sigma_{\delta^{i}} = 0.0175$ rad based on the information obtained from the real-time experiments. Thus, the following weighting matrices $V_y$, $V_u$ and $V_s$ have been used in NMHE:
\begin{eqnarray}\label{weightingmatricesVyVu}
V_{y} & = & diag(\sigma_{x^t},\sigma_{y^t},\sigma_{x^i},\sigma_{y^i}, \sigma_{\beta} ,\sigma_{v})^T \nonumber \\
        & = & diag(0.03,0.03,0.03,0.03,0.0175,0.01)^T \nonumber \\
V_{u} & = & diag(\sigma_{\delta^{t}},\sigma_{\delta^{i}})^T \nonumber \\
        & = & diag(0.0175,0.0175)^T \nonumber \\
V_{s} & = & diag(x^t, y^t, \theta, x^i, y^i, \psi, \mu, \kappa, \eta, \beta, v)^T \nonumber \\
      & = & diag(10.0, 10.0, 0.1, 10.0, 10.0, 0.1, 0.25, 0.25, 0.25, 0.1745, 0.1)^T
\end{eqnarray}

\subsection{Solution Methods}

The optimization problems in NMPC \eqref{nmpc} and in NMHE \eqref{mhe} are similar to each other, and it can be clearly seen that the same solution method can be applied for both NMPC and NMHE \cite{TomKraus}. In this paper, the multiple shooting method has been used in a fusion with a generalized Gauss-Newton method. This method is a version of the classical Newton method and developed for least-squares problems. The main advantage of it is that it does not require the second derivatives which are difficult to compute. On the other hand, the disadvantage is that since it is an iterative method and it is hard to predict the necessary number of the iteration in order to reach a desired accuracy. Although the number of iterations cannot be determined in advance, a simple solution was proposed in \cite{Diehl2} in which the number of Gauss-Newton iterations is limited to $1$. Meanwhile, each optimization problem is initialized with the output of the previous one. This proposed method gives similar results with less computational burden when compared to the classical one with the advantage of minimum feedback delay.

The \emph{ACADO} code generation tool, an open source software package for optimization problems \cite{ACADO}, has been used to solve the constrained nonlinear optimization problems in the NMPC and NMHE. First, this software generates C-code, which is then converted into a .dll file to be used in \emph{LabVIEW}. Detailed information on the \emph{ACADO} code generation tool can be found in \cite{home,Houska2011a,ACADO}.

%%%%%%%%%%%%%%%%%%%%%%%%%%%%%%%%%%%%%%%%%%%%%%%%%%%%%%%%%%%%%%%%%%%%%%%%%%%%%%%%%%%%%%%%%%%%%%%%%%%%%%%%%%%%%%%%%%%%%%%%%%%%%%%%%%%%%%%%%%%%%%%%%%%%%%%%%%%%%%%%
\section{Experimental Set-Up Description and Real Time Results} \label{realtime}
\subsection{Experimental Set up Description}
\begin{figure}[b!]
\centering
  \includegraphics[width=3.5in]{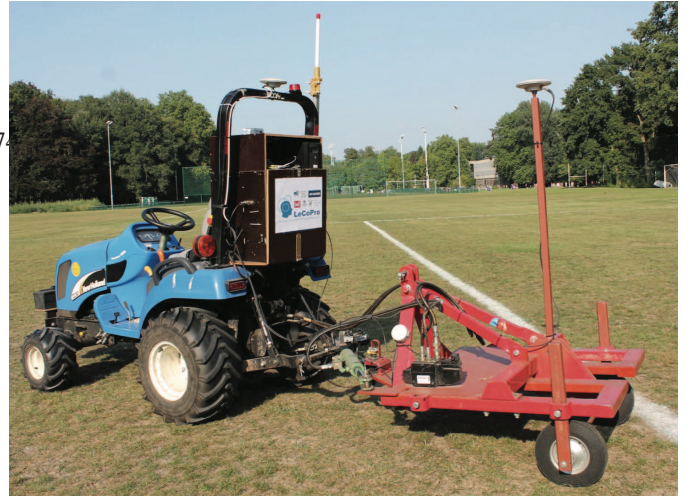}\\
  \caption{The tractor-trailer system}
  \label{tractor1}
\end{figure}

The real-time set-up which is a small agricultural tractor-trailer system is illustrated in Fig. \ref{tractor1}. Two GPS antennas have been mounted straight up the center of the tractor rear axle and the center of the trailer to provide highly accurate positional information. They are connected to a Septentrio AsteRx2eH RTK-DGPS receiver (Septentrio Satellite Navigation NV, Belgium) with a specified position accuracy of 2cm at a 5-Hz sampling frequency. The Flepos network supplies the RTK correction signals via internet by using a \emph{Digi Connect WAN 3G} modem.

\begin{figure}[t!]
\centering
\includegraphics[width=3.5in]{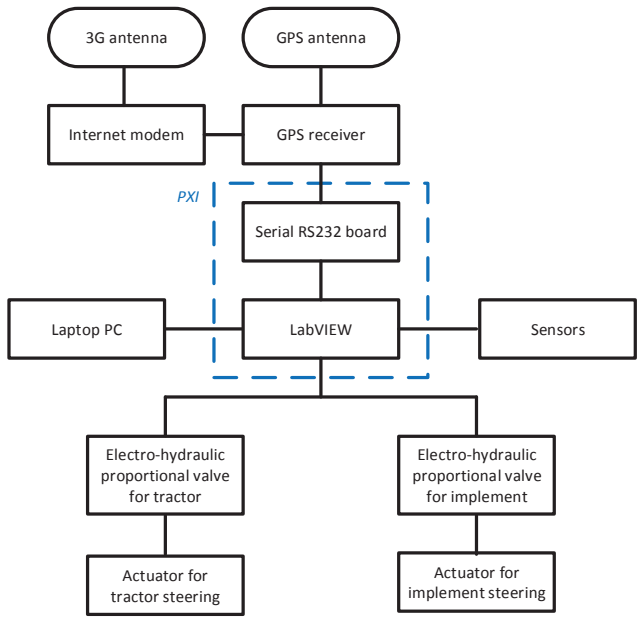}\\
\caption{Block diagram of hardware}
\label{blockdiagramhardware2}
\end{figure}

The block diagram of the hardware is shown in Fig. \ref{blockdiagramhardware2}. The GPS receiver and the internet modem are connected to a real time operating system (PXI platform, National Instrument Corporation, USA) via a RS232 serial communication. The PXI system gathers the steering angles, the GPS data and controls the tractor-trailer system by sending messages to the electro-mechanical actuator for the gas pedal and electro-hydraulic actuator for the steering mechanism. A laptop connected via WIFI to the PXI system functions as the user interface of the autonomous tractor. The control algorithms are implemented in $LabVIEW^{TM}$ (v2011, National Instruments, Austin, TX). They are executed in real time on the PXI and updated at a rate of 5-Hz.

The CeNMPC calculates the desired steering angle for the front wheels of the tractor and the actuator of the trailer, and two low level controllers, PI controllers in our case, are used to control the steering mechanisms. While the angle of front wheels of the tractor is measured using a potentiometer mounted on the front axle yielding a steering angle resolution of $1^{\circ}$, the angle between the trailer and the drawbar is measured by using an inductive sensor which has $1^{\circ}$ precision.

The speed of the tractor is controlled by using an electro-mechanic valve. There are two PID type controllers in the speed control system. The PID controllers in the outer closed-loop and inner closed-loop are generating the desired pedal position with respect to the speed of the tractor and volt for the electro-mechanic valve with respect to the pedal position, respectively.

\subsection{Real-time Results}
During the real-time experiments, an 8-shaped trajectory, which consists of two straight lines and two smooth curves, is tracked. Such a trajectory allows us to investigate the performance of the designed controller for both straight line and curvilinear geometries. The curvature of the smooth curves is equal to 0.1. (The curvature of a circle is the inverse of its radius). It is to be noted that the field is not flat, but it contains bumps and holes.

In this paper, a space-based trajectory is preferred instead of using a time-based trajectory. It is to be noted that we do not force the vehicle to be at a specific point on the trajectory  at a specific time instant in space-based trajectories. The reference generation method in this paper is as follows: As soon as the tractor starts off-track, first, it quickly calculates the closest point on the space-based trajectory, then it determines its desired point. The desired point is a fixed forward distance from the closest point on the trajectory at every specific time instant. While the selection of a big distance from the closest point on the trajectory  results in a steady-state error on the trajectory following, a small distance would result in oscillations during the trajectory tracking. The main goal of the reference generation algorithm for the tractor is to minimize the steady-state tracking error with minimal oscillation around the desired trajectory. The optimum distance, which has been determined by trial-and-error method in this paper, was found to be $1.6$ meters ahead from the front axle of the tractor. It is to be noted that this value depends on the longitudinal velocity of the tractor. The selected value for this study is for $1$ $m/s$ longitudinal velocity.

\begin{figure}[b!]
\centering
\includegraphics[width=3.5in]{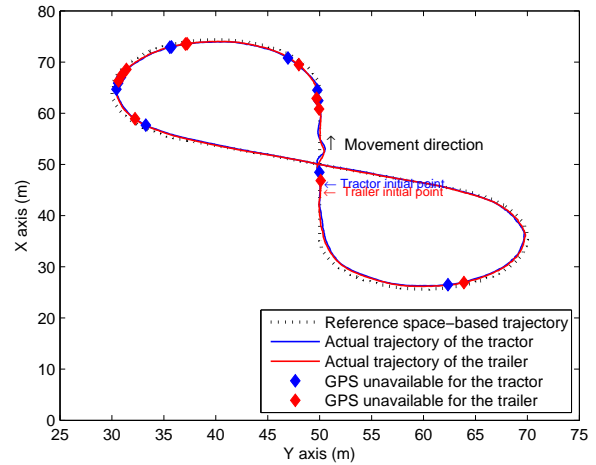}\\
\caption{Reference trajectory, vehicle actual trajectory and unavailable GPS signals on the trajectory}
\label{traj}
\end{figure}

As can be seen from Fig. \ref{traj}, the autonomous tractor is capable to stay on-track after a finite time. One of the most serious problems of such outdoor autonomous vehicles using GPS is the missing data values from the satellites. As can be seen in Fig.  \ref{traj}, there were 11 missing data points within 871 data points during the experiment. The mentioned problem is an inevitable drawback of using RTK-GPS, and the control algorithm must be robust enough to cope with this. Theoretically, NMHE cannot handle the missing data problem after having certain number of consecutive outliers. The maximum number of outliers allowed for a robust estimation depends on the selected estimation horizon.

\begin{figure}[t!]
\centering
\includegraphics[width=3.5in]{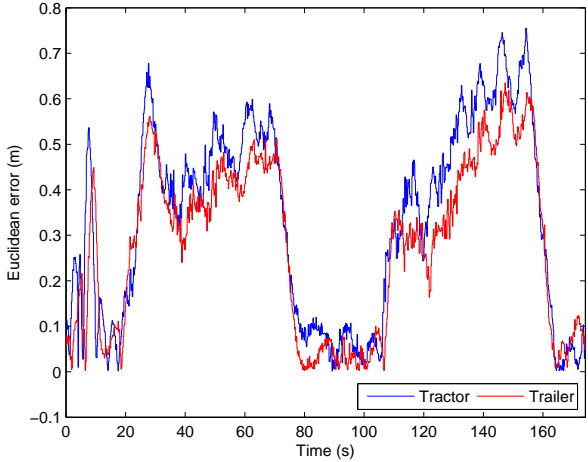}\\
\caption{Euclidean error to the space-based reference trajectory}
\label{error}
\end{figure}

The Euclidean error to the space-based reference trajectory for both the tractor and the trailer has been shown in Fig. \ref{error}. The Euclidean error values of the tractor and the trailer for the straight lines are approximately 0.0644m and 0.0361m, respectively. Besides, the Euclidean error values of the tractor and the trailer for the curved lines are approximately 0.4978m and 0.4152m, respectively. The reason for having big error values for curvilinear trajectories is because we have selected a space-based trajectory instead of a time-based one. It is to be noted that this problem is an inevitable drawback of having a space-based trajectory. However, in agricultural field operations, mostly straight line trajectories are used. Although CeNMPC calculates the proper outputs for $\delta^{i}$ at $RJ^2$, the error correction for the trailer is limited due to the limited length of the drawbar between the tractor and the trailer (equal to 20cm in this set up). In \cite{TomKraus}, while the author uses a time-based trajectory, a space-based trajectory is preferred in this study. It has been observed that the controller in this investigation outperforms the results in \cite{TomKraus}.

\begin{figure}[t!]
\centering
\includegraphics[width=3.5in]{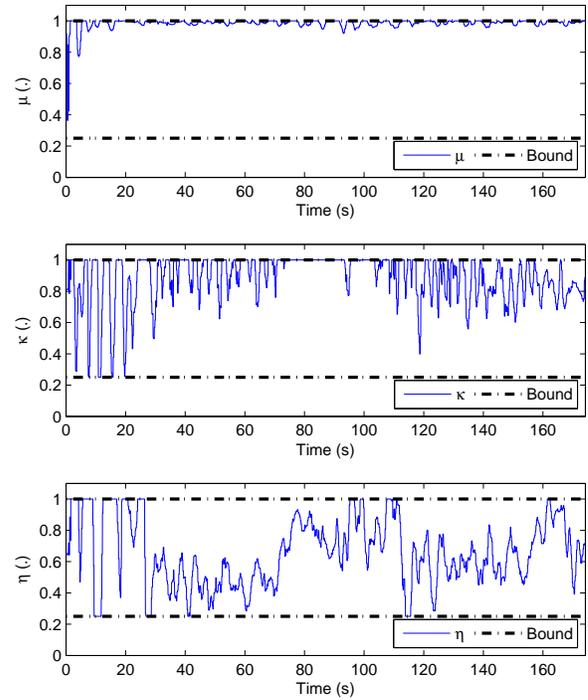}\\
\caption{Tractor longitudinal slip coefficient $(\mu)$, tractor ($\kappa$) and trailer ($\eta$) side slip coefficients}
\label{slips}
\end{figure}

In Fig. \ref{slips}, the NMHE parameter estimation performance for the slip coefficients is presented. As can be seen from this figure, the estimated parameter values are within the constraints in \eqref{constraints2}. The variations in these slip parameters not only stand for the real slip values in the system but also for the lack of modeling. The kinematic model used in this paper in not capable of representing all the system dynamics and interactions in the real-time system.

\begin{figure}[h!]
\centering
\includegraphics[width=3.5in]{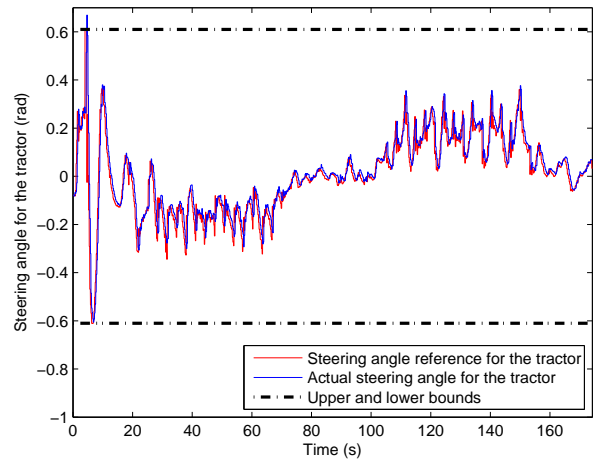}\\
\caption{Reference and actual steering angle}
\label{steering}
\end{figure}

\begin{figure}[h!]
\centering
\includegraphics[width=3.5in]{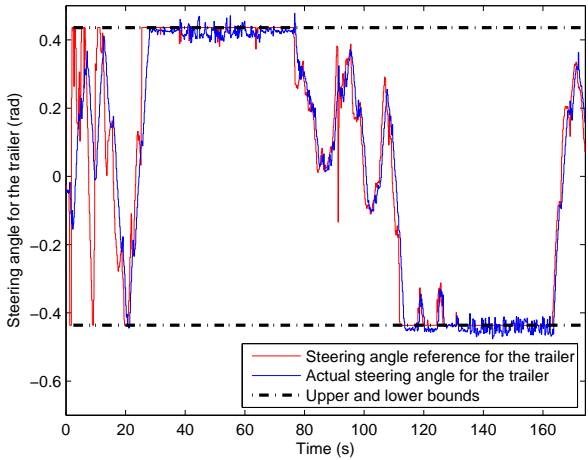}\\
\caption{Reference and actual hitch point angle}
\label{hitch}
\end{figure}

In Figs. \ref{steering}-\ref{hitch}, the outputs, the steering angle reference for the tractor ($\delta^{t}$) and the steering angle reference for the trailer ($\delta^{i}$), of the CeNMPC are shown. As can be seen from this figure, the performance of the low level controllers is sufficient to track the reference signals generated by the CeNMPC. However, it should be noted from Fig. \ref{hitch} that the steering angle reference for the trailer hits the constraint of the maximum achievable angle. This can be explained by the aforementioned short drawbar length which makes that the lateral deviation of the trailer is not sufficient to correct for the tracking error even at the maximum steering angle.

Moreover, as can be seen from Figs. \ref{steering} and \ref{hitch}, the reference steering angles generated by CeNMPC for the tractor and trailer are within the upper and lower bounds. However, the steering angles are controlled by conventional PI controllers. The actual values of the steering angles have been violated around 1 degree due to the transient response or the noise on the measurements.


The execution times for CeNMPC and NMHE are summarized in Table \ref{mpcmheperformance}. During the real-time experiments, a real-time controller supplied with a 2.26 GHz Intel Core 2 Quad Q9100 quad-core processor (NI PXI-8110, National Instruments, Austin, TX, USA) has been used. The NMHE and the NMPC routine have been assigned to one processor on which the other runs in serial. As can be seen from this table, the CeNMPC-MHE framework is able to give overall computational times less than $7.2$ms in the worst case when \emph{ACADO} code generation tool is used.

\begin{table}
\centering
\caption{Execution times of the CeNMPC and the NMHE.}\label{mpcmheperformance}
\begin{tabular}{lccc}
  \hline
   &  Minimum & Average &  Maximum (ms)\\
   &  (ms) &   (ms) &  (ms) \\
    \hline
  CeNMPC CPU time & &  &  \\
  Preparation & 6.5462 & 6.6632 & 6.9260 \\
  Feedback & 0.0521 &  0.1345 &  0.3140 \\
  Overall & 6.5983 &  6.7977&  7.2400 \\
  \hline
    \hline
  NMHE CPU time &  \\
  Preparation & 4.000 & 4.1002 & 4.500\\
  Estimation & 0.3276 & 0.5813 & 1.4432\\
  Overall & 4.3276 & 4.6813 & 5.9432 \\
  \hline
\end{tabular}
\end{table}

\section{Conclusions and Future Research} \label{Conc}
\subsection{Conclusions}
A fast CeNMPC-MHE framework has been elaborated for the control of an autonomous tractor-trailer system. The forward and side slip parameters and the tractor and trailer yaw angles, which cannot be measured directly, but are essential for the trajectory tracking, have been estimated with NMHE. The parameter estimation results show the efficiency of the NMHE. The experimental results show that the proposed CeNMPC-MHE framework is able to control the tractor-trailer system with a reasonable accuracy. The Ecludian error to the straight line trajectory is around 0.0644m and 0.0361m for the tractor and the trailer, respectively. Although CeNMPC is more computationally intensive than decentralized control approaches, the optimizations could be performed sufficiently fast thanks to the use of the ACADO toolkit.

\subsection{Future research}
Although the designed CeMPC-MHE framework is based-on the kinematic model of the tractor-trailer system, due to the fact that ACADO toolkit gives feedback times at around 6-7 ms, it is amenable to extend the CeNMPC-MHE framework to use a dynamic model to have a better control performance in transient responses of the system.

\section*{Acknowledgment}
This work has been carried out within the framework of the LeCoPro project (IWT-SBO 80032) funded by the Institute for the Promotion of Innovation through Science and Technology in Flanders (IWT-Vlaanderen). We would like to thank Mr. Soner Akpinar for his technical support for the preparation of the experimental set up.

\ifCLASSOPTIONcaptionsoff
  \newpage
\fi

\bibliography{cenmpc_bib}
\bibliographystyle{IEEEtran}

\begin{IEEEbiography}[{\includegraphics[width=1in,height=1.25in,clip,keepaspectratio]{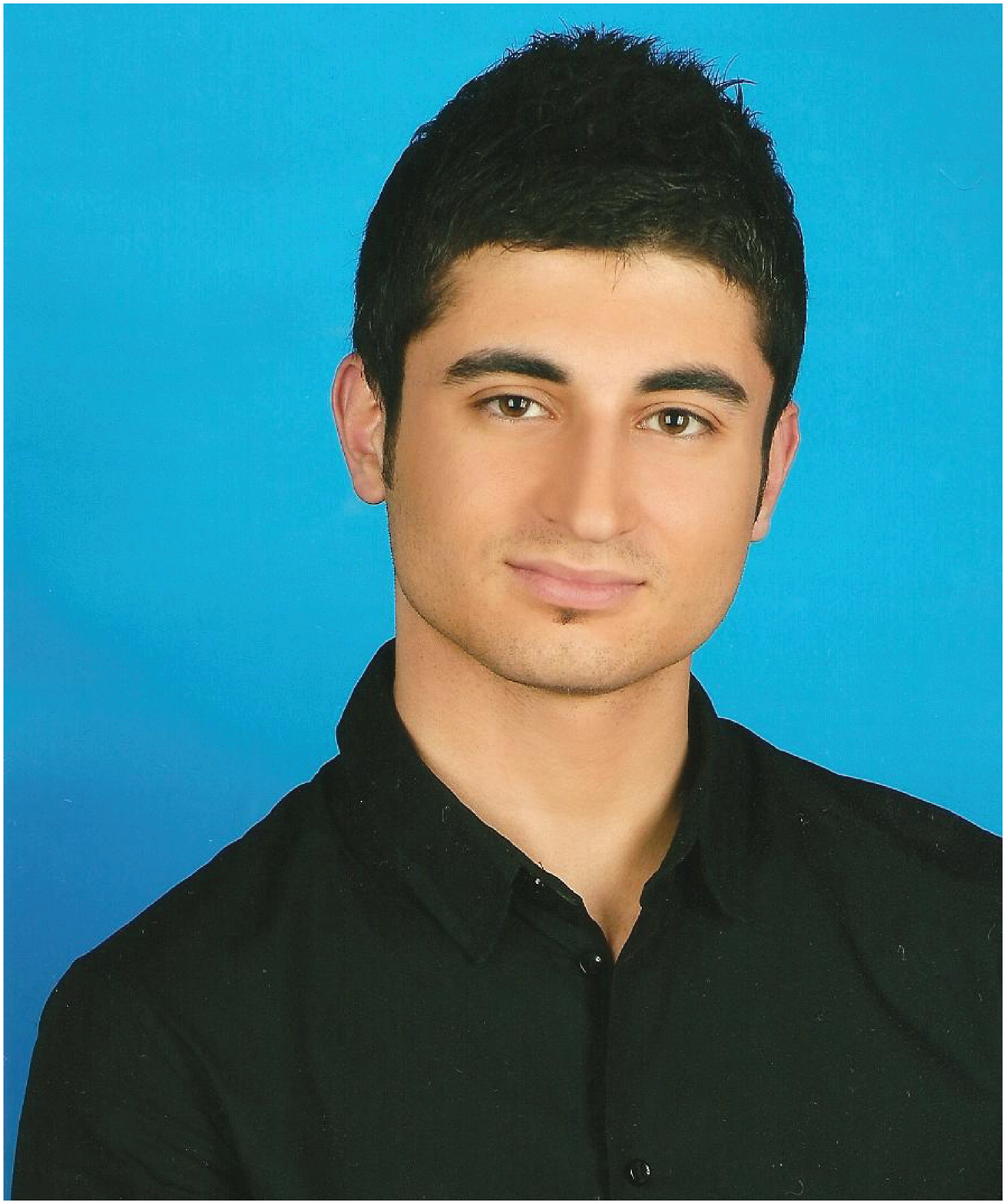}}]{Erkan Kayacan} (S\textquoteright 12) was born in Istanbul, Turkey, on April 17, 1985. He received the B.Sc. and the M.Sc. degrees in mechanical engineering from Istanbul Technical University, Istanbul, in 2008 and 2010, respectively. He is a PhD student and research assistant at University of Leuven (KU Leuven) in the division of mechatronics, biostatistics and sensors (MeBioS). His research interests include model predictive control, moving horizon estimation, distributed and decentralized control, intelligent control, vehicle dynamics and mechatronics.

He was active in the IEEE CIS GOLD Subcommittee and the IEEE Webinars subcommittee.
\end{IEEEbiography}

\begin{IEEEbiography}[{\includegraphics[width=1in,height=1.25in,clip,keepaspectratio]{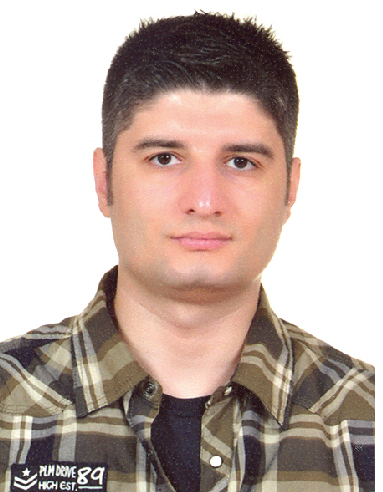}}]{Erdal Kayacan} (S\textquoteright 06-SM\textquoteright 12)  was born in Istanbul, Turkey on January 7, 1980. He received a B.Sc. degree in electrical engineering from in 2003 from Istanbul Technical University in Istanbul, Turkey as well as a M.Sc. degree in systems and control engineering in 2006 from Bogazici University in Istanbul, Turkey. In September 2011, he received a Ph.D. degree in electrical and electronic engineering at Bogazici University in Istanbul, Turkey. After finishing his post-doctoral research in KU Leuven at the division of mechatronics, biostatistics and sensors (MeBioS), he is currently pursuing his research in Nanyang Technological University at the School of Mechanical and Aerospace Engineering as an assistant professor. His research areas are unmanned aerial vehicles, robotics, mechatronics, soft computing methods, sliding mode control and model predictive control.

Dr. Kayacan has been serving as an editor in Journal on Automation and Control Engineering (JACE) and editorial advisory board in Grey Systems Theory and Application.
\end{IEEEbiography}

\begin{IEEEbiography}[{\includegraphics[width=1in,height=1.25in,clip,keepaspectratio]{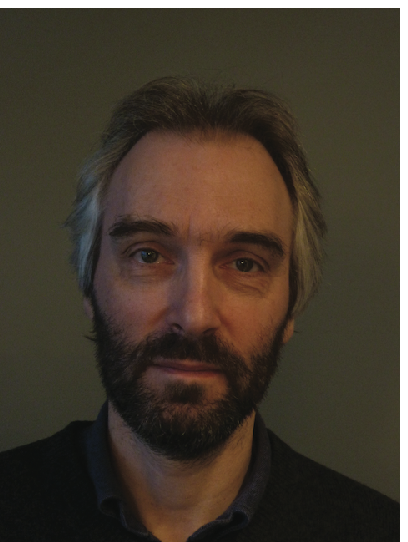}}]{Herman Ramon} graduated as an agricultural engineer from Gent University. In 1993 he obtained a Ph.D. in applied biological sciences at the Katholieke Universiteit Leuven. He is currently Professor at the Faculty of Agricultural and Applied Biological Sciences of the Katholieke Universiteit Leuven, lecturing on agricultural machinery and mechatronic systems for agricultural machinery. He has a strong research interest in precision technologies and advanced mechatronic systems for processes involved in the production chain of food and nonfood materials, from the field to the end user. He is author or co-author of more than 90 papers.
\end{IEEEbiography}

\begin{IEEEbiography}[{\includegraphics[width=1in,height=1.25in,clip,keepaspectratio]{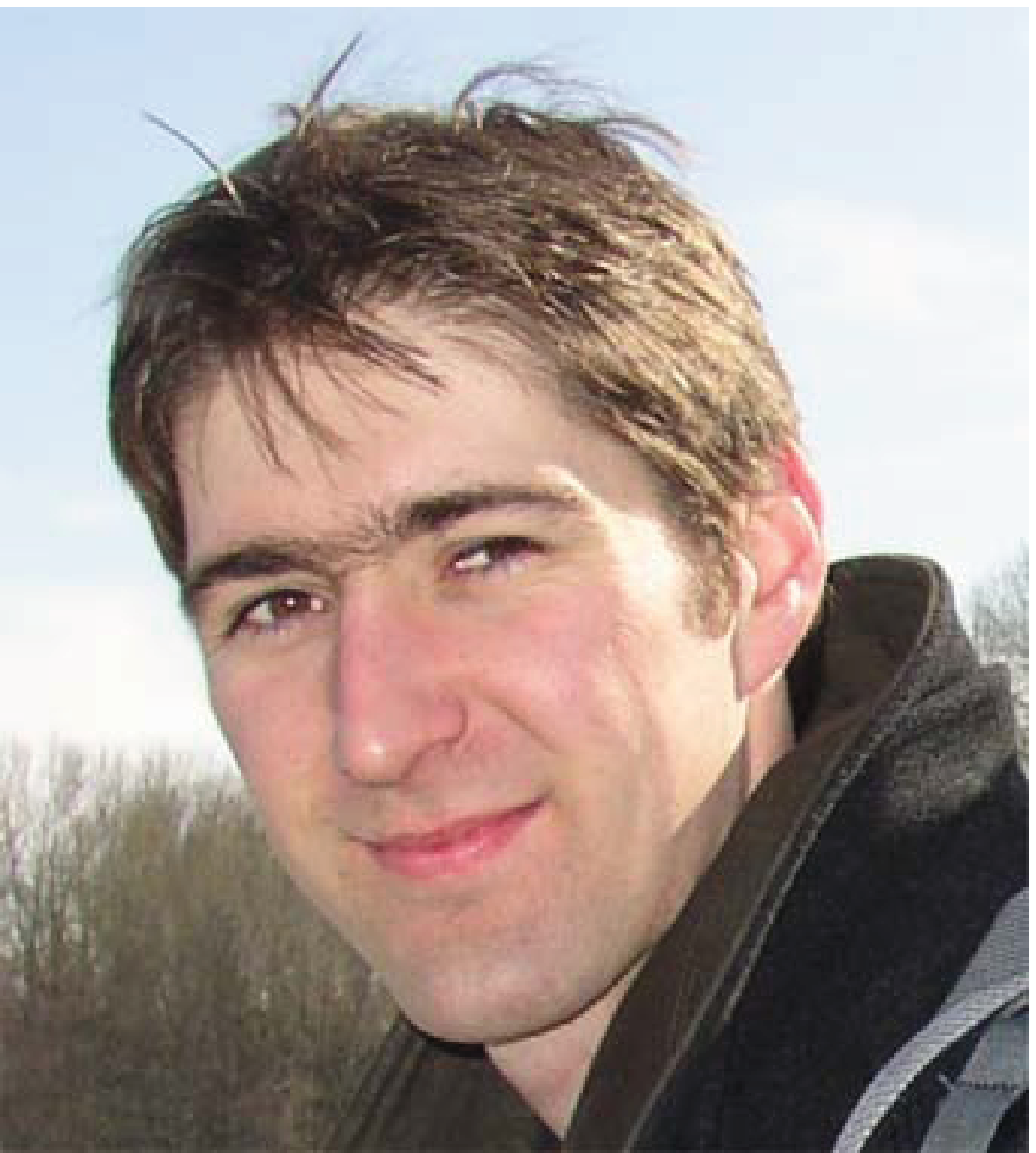}}]{Wouter Saeys} is currently Assistant Professor in Biosystems Engineering at the Department of Biosystems of the University of Leuven in Belgium. He obtained his Ph.D. at the same institute and was a visiting postdoc at the School for Chemical Engineering and Advanced Materials of the University of Newcastle upon Tyne, UK and at the Norwegian Food Research Institute - Nofima Mat in Norway. His main research interests are optical sensing, process monitoring and control with applications in food and agriculture. He is author of 50 articles (ISI) and member of the editorial board of Biosystems Engineering.
\end{IEEEbiography}

\clearpage

\end{document}